\documentclass[11pt]{article}

\usepackage[utf8]{inputenc}

\usepackage{amsthm,amsmath,amssymb}
\RequirePackage[numbers]{natbib}
\RequirePackage[colorlinks,citecolor=blue,urlcolor=blue]{hyperref}
\usepackage{array}
\usepackage{algorithm}
\usepackage{algorithmic}
\usepackage{color}
\usepackage{times}
\usepackage{graphicx}

\newtheorem{theorem}{Theorem}

\newtheorem{lemma}{Lemma}
\newtheorem{proposition}{Proposition}

\def\bo{\mathring{\beta}}
\def\b*{\mathring{\beta}}

\def\bmin{\mathring{\beta}_{min}}

\def\Ex{\mathbb E}
\def\ml{\dot{\ell}}

\def\hbeta{\hat{\beta}}

\def\mA{\mathcal{A}}

\def\J{M}
\def\T{M_ {\mathring{\beta}}}

\def\cone{C}

\title{Improving Group Lasso for high-dimensional categorical data}

\author{Szymon Nowakowski \footnote{Faculty of Physics, University of Warsaw, Pasteura 5, 02-093 Warsaw, Poland}
\footnote{Institute of Applied Mathematics and Mechanics,
University of Warsaw, Banacha 2, 02-097 Warsaw, Poland , sd.nowakowski2@uw.edu.pl },
Piotr Pokarowski \footnote{Institute of Applied Mathematics and Mechanics,
University of Warsaw, Banacha 2, 02-097 Warsaw, Poland, 
pokar@mimuw.edu.pl }, \\
Wojciech Rejchel 
\footnote{Faculty of Mathematics and Computer Science,
Nicolaus Copernicus University,  Chopina 12/18, 87-100, Toru\'n, Poland, 
wrejchel@gmail.com } 
, Agnieszka So{\l}tys \footnote{Institute of Applied Mathematics and Mechanics,
University of Warsaw, Banacha 2, 02-097 Warsaw, Poland, 
agnieszkaprochenka@gmail.com }}

\date{}

\begin{document}

\maketitle

\begin{abstract}
 Sparse modelling or model selection with categorical data is challenging even for a moderate number of variables, because one parameter is roughly needed to encode one category or level. The Group Lasso is a well known efficient algorithm for selection continuous or categorical variables, but all estimates related to a selected factor usually differ. Therefore, a fitted model may not be sparse, which makes the model interpretation difficult. To obtain a sparse solution of the Group Lasso we propose the following two-step procedure: first, we reduce data dimensionality using the Group Lasso; then to choose the final model we use an information criterion on a small family of models prepared by clustering levels of individual factors. We investigate selection correctness of the algorithm in a sparse high-dimensional scenario. We also test our method on synthetic as well as real datasets and show that it performs better than the state of the art algorithms with respect to the prediction accuracy or model dimension.
\end{abstract}

\section{Introduction}
\label{sec:intro}

Data sets containing categorical variables (factors) are common in statistics and machine learning. The analysis of such data sets is much more challenging than for those having only numerical variables.  There are two main reasons of that: the first one relates to the fact that  a factor with $k$ levels is usually encoded as $k-1$ dummy variables, so $k-1$ parameters are needed to learn (there is only one such parameter for a numerical predictor).
Therefore, dimensionality reduction is crucial, which becomes the second reason. Namely, 
dimensionality reduction is  much more involved for factors than for numerical precitors. 
Indeed, for the latter we have only two possibilities (leave or delete a predictor). However, for a categorical predictor we can either delete it or merge some of its levels. In the latter case a number of possibilities grows very quickly. For instance, consider the factor corresponding to a continent that a person (client, patient etc.) lives  or a company (university etc.) is located. This factor has 6 levels (Antarctica is not considered), which gives 203 possibilities (they are usually called {\it partitions}). 

Due to that, it is really difficult to  develope efficient algorithms for categorical data and investigate their statistical properties. It is especially apparent in the case that a number of factors and/or  numbers of their levels are large. Thus, this topic has not been intensively studied and the corresponding literature has been relatively modest. However, categorical data are so common that it must have changed. Indeed, we have found many papers investigating categorical data from last few years, among others \citet{garcia2021variable, pauger2019bayesian,rabin_rosset_jasa, rabin_rosset_arxiv, simchoni_rosset21,  stokell2021}. 

In the current paper we consider high-dimensional selection and sparse prediction, where a number of active variables is significantly smaller than  learning sample size $n$ and  number of all variables $p$ greatly exceeds $n.$ Popular methods fitting sparse predictive models with high-dimensional data do not merge levels of factors: the Lasso \citep{Tibshirani96} treats dummy variables as separate, binary predictors, the Group Lasso \citep{groupLasso} can only leave or delete a whole factor and  the Sparse Group Lasso \citep{simon2013sparse} additionally removes levels of selected factors. 
The Fused Lasso \citep{fused2005} can merge levels, but only in a simplified case that variables are ordered.  These methods  significantly reduce a number of parameters and select
 variables, which may be an input for further, interpretable dimension reduction techniques. However, they do not realize {\it partition selection}. It means that they cannot choose models consisting of subsets of numerical variables and partitions of levels of factors.

In the mainstream research on the Lasso-type algorithms, the CAS-ANOVA method \citep{bondell2009} fits sparse linear models with fusion of factor levels using the $l_1$ penalty imposed on differences between parameters corresponding to levels of a factor. The implementation of CAS-ANOVA is provided in \citet{gertheiss2010sparse, oelker14}. An alternative to the penalization is a greedy search algorithm from \citet{MajK2015}. As we have already mentioned,  a growing interest in partition selection has been noticed recently. \citet{pauger2019bayesian} introduced a Bayesian method for linear models based on a prior inducing fusion of levels. Another approach trying to solve the problem from the Bayesian perspective is considered in \citet{garcia2021variable}. The frequentist method using linear mixed models was presented in \citet{simchoni_rosset21}, where factors were treated as random effects. A partition selection algorithm called SCOPE, which is based on a regularized likelihood, can be found in \citet{stokell2021}. This procedure uses a minimax concave penalty on differences between consecutive, sorted estimators of coefficients for levels of a factor. 
Finally, trees-based algorithms are applied to categorical data in \citet{rabin_rosset_arxiv}.

Let us note that all methods from the above paragraph, which make partition selection are restricted to a classical scenario $p<n$, except SCOPE and DMR, which in its new implementation is based on variables screened by the Group Lasso \citep{DMRnet}. 
In this paper, we present an improved as well as simplified version of the DMR algorithm, called PDMR (Plain DMR), which is much simpler for a theoretical research and also works better for numerical experiments, in particular:

1. We propose the following two-step procedure PDMR: first, we reduce data dimensionality using the Group Lasso; then to choose the final model we use an information criterion on a small family of models prepared by clustering levels of individual factors based on the Group Lasso estimator. We prove in Theorem \ref{main_th} that  under weak conditions  that PDMR returns a sparse linear model containing the true model, even if $p >> n$. The proof is based on a new bound of a number of partitions by generalized Poisson moments.
 It is worth to note that so far there are no theoretical results regarding the correctness of the DMR selection for high-dimensional data, while for SCOPE, a weaker property than selection consistency was proved in \citet{stokell2021}. Our result is also weaker than  selection consistency, but 
it relates directly to an output of our algorithm, while results from \citet[Theorem 6]{stokell2021} concerns one of 
blockwise optima of their objective function, so an output of their algorithm need not possess this property.

2. To investigate theoretical properties of PDMR we require probabilistic inequalities for an estimation error of Group Lasso. Such results exist in the literature, but when applied to partition selection they give sub-optimal results. Therefore, in Subsection \ref{subsec:groupL}   
we prove a novel bound for  the $l_{\infty}$ estimation error of the weighted Group Lasso.

3. As a by-product of results from Subsection \ref{subsec:groupL} we obtain  optimal weights for the Group Lasso, which are different from those recommended by the authors of this  method \citep{groupLasso}. Possibly, the new weights can improve asymptotics of the Group Lasso in the general scenario (not necessarily orthogonal, which is considered in Subsection \ref{subsec:groupL})  and its practical performance as well.

4. In theoretical considerations the Lasso-type algorithms are defined for one penalty and return one coefficient estimator. However, practical implementations usually use nets of data-driven penalties and return lists of estimators. Our next contribution is  an analogous implementation of the PDMR algorithm. Next, in numerical experiments on high-dimension data, we compare PDMR, DMR and SCOPE. 
We show on a set of linear models (in wider scenarios than in \citet{stokell2021}) that PDMR give results comparable or better in terms of RMSE  than SCOPE  and  DMR. While on a set of several real data sets we observe that the results of these three algorithms are similar. Moreover, PDMR and DMR are several dozen times faster than SCOPE.

In the rest of this paper we describe the considered models and the PDMR algorithm. We also present mathematical propositions with proofs, which describe properties of our method. Finally, we compare PDMR to other methods  in numerical experiments. All proofs are relegated to supplementary materials.

\section{Linear models and the algorithm}
\label{sec:model_alg}

We consider independent data $(y_1,x_{1.}),(y_2,x_{2.}),\ldots,(y_n,x_{n.})$, 
where $y_i \in \mathbb{R}$ is a response variable and $x_{i.} \in\mathbb{R}^p$ is a vector of predictors. Every vector of predictors $x_{i.}$ can consist of continuous predictors as well as categorical predictors. 
We arrange them in the following way
$x_{i.}=(x_{i1}^T,x_{i2}^T,\ldots,x_{ir}^T).^T
$
Suppose that   $x_{ik}$ corresponds to a categorical predictor (factor) for some $k \in \{1,\ldots,r\}.$ A set of  levels of this factor is given by $\{0,1,2,\ldots,p_k\}$ and $x_{ik} \in \{0,1\}^{p_k}$ is a dummy vector corresponding to the $k$-th predictor of the $i$-th object in a data set. So, a reference level (say, the zero level) is not included in $x_{ik}.$
The only exception relates to the first factor, whose  reference level  is  contained in $x_{i1}.$ This special level plays a role of an intercept. If necessary, we can rearrange vectors of predictors in such a way to have the first factor with $k=1.$
If $x_{ik}$ corresponds to a continuous predictor, then simply $x_{ik} \in \mathbb{R}^{p_k}$ and $p_k=1.$
Therefore, a dimension of $x_{i.}$ is   $p=1+\sum_{k=1}^r p_k.$ Finally, let $X=[x_{1.},\ldots,x_{n.}]^T$ be a $n\times p$ design matrix.

We consider a linear model 
\begin{equation}
\label{inverseLink}
 y_i=x_{i.}^T\mathring{\beta} + \varepsilon_i \quad \text{for} \quad i=1,2,\dots,n.
\end{equation}
Coordinates of $\bo$ correspond to coordinates of a vector of predictors, that is
$
\bo=(\bo _1^T, \bo _2^T, \ldots, \bo _r^T
)^T,
$
where $\bo _1 = (\bo _{0,1}, \bo _{1,1}, \ldots,
\bo _{p_1,1} )^T\in \mathbb{R}^{p_1+1}$ and  $\bo _k = (\bo _{1,k}, \bo _{2,k}, \bo _{3,k}, \ldots,
\bo _{p_k,k} )^T\in \mathbb{R}^{p_k}$ for  $k=2,\ldots, r.$ 
Moreover, we suppose that noise variables $\varepsilon_i$ have a {\it subgaussian distribution}
with the same  number $\sigma>0$, that is for $i=1,2,\dots,n$ and $u \in \mathbb{R}$ we have
\begin{equation}
\label{subgauss}
\mathbb{E}\exp(u\varepsilon_i) \leq \exp(\sigma^2u^2/2).
\end{equation}
The main examples of subgaussian noise variables are normal variables or those having bounded supports.

\subsection{Notations}
\label{sec_notations}

Let $W_1=diag(w_{0,1},w_{1,1},\ldots, w_{p_1,1})$ and $W_k=diag(w_{1,k},\ldots, w_{p_k,k}),k=2,\ldots,r$ be diagonal nonrandom matrices with positive entries. Besides, $W=diag(W_1,\ldots, W_r)$ is a $p \times p$ diagonal matrix with matrices $W_k$ on the diagonal. Next, for $\beta \in \mathbb{R}^p$ and $q\geq 1$  
let $|\beta|_q = (\sum_{j=1}^p |\beta_j|^q)^{1/q}$ be the $\ell_q$ norm of $\beta$. The only  exception is  the $\ell _2$ norm, for which we use the special notation $||\beta||.$

A feasible model is defined as a sequence $M=(P_1,P_2,\ldots,P_r).$ If the $k$-th predictor is a factor, then $P_k$ is a particular partition of its levels. If the $k$-th predictor is continuous, then $P_k \in \{\emptyset,\{k\}\}.$
To make the notation coherent and concise we {\it artificially} augment each $\beta \in \mathbb{R}^p$ by 
 $\beta_{0,k}=0, k=2, \ldots,r.$
Notice that every $\beta $ determines a model $M_\beta$ as follows: if the $k$-th predictor is a factor, then partition  $P_k$ depends on the fact, whether equalities  $\beta_{j_1,k}=\beta_{j_2,k}, j_1 \neq j_2$ are satisfied, which corresponds to merging levels $j_1$ and $j_2$.  If the $k$-th predictor is continuous, then $P_k=\{k\}$ when $\beta_k \neq 0$ and $P_k=\emptyset$ otherwise.

In the following we consider $k \in \{1,\ldots, r\}$ and for $k=2,\ldots,r$ we have  $j \in \{1,\ldots,p_k\}, $ while for $k=1$ we have $j \in \{0,\ldots,p_1\}. $ Let  $x_{j,k}$ be a column of $X$ corresponding to the $j$-th level ofthe  $k$-th factor. The additional notations are $x_M=\max_{j,k} \; ||x_{j,k}||, x_m=\min_{j,k} \; ||x_{j,k}||, x_W=\max_{j,k}\; ||x_{j,k}||/w_{j,k}.$
Finally, we define a crucial object $\Delta= \min\limits_{1\leq k\leq r} \;\;\min\limits _{0\leq j_1,j_2\leq p_k: \bo _{j_1,k} \neq \bo _{j_2,k}} |\bo _{j_1,k} - \bo _{j_2,k}|.$ It is involved in an analog of the {\it beta-min} condition, which plays an important role when variable selection with  continuous predictors is considered \citep{BuhlmannGeer11, YeZhang10}.

\subsection{The algorithm}

To simplify notations (and without loss of generality), we suppose that all considered predictors are categorical. For estimation of  $\mathring{\beta}$ we consider a  quadratic loss function
\begin{equation}
\label{loss}
 \ell(\beta)=\sum_{i=1}^n [(x_{i.}^T\beta)^2/2-y_ix_{i.}^T\beta]
\end{equation}
as in maximum likelihood estimation.
It is easy to see that  $\dot \ell(\beta) =\sum_{i=1}^n (x_{i.}^T\beta-y_i) x_{i.}$, where $\dot \ell$ denotes a derivative of $\ell .$
Besides,  $\dot \ell(\mathring{\beta}) = -X^T\varepsilon$ for $\varepsilon=(\varepsilon_1,\ldots,\varepsilon_n)^T$. Next,
for $k=1,\ldots, r$ partial derivatives of $\ell (\beta)$ corresponding to coordinates of $\beta_k$ are denoted by $\dot \ell _k(\beta).$ 

We present PDMR, which consists of two steps: \\
{\bf (1)  Screening}: we compute the weighted Group Lasso $$\hat{\beta} =\arg \min_{\beta} \ell (\beta) + \lambda 
\sum_{k=1}^r ||W_k \beta_k|| ,
  $$
where diagonal elements $(W_k)_{jj}=||x_{j,k}||$ play roles of weights. Such a choice of weights is explained in Proposition \ref{opt_w},\\
{\bf (2)  Selection}: this step is divided into three parts:\\
{\bf (2a)  construction of nested family of models $\mathcal{M}$:} let $\hat S=\{1 \leq k\leq r: \hat \beta_k \neq 0\}$ and $\hat \beta_{0,k}=0$ for $ k \in \hat S \setminus \{1\}.$ So, $\hat S$ is a set of factors which are not discarded by the Group Lasso. 
For each $k \in \hat S$  we  separately perform
 complete linkage clustering of levels of those factors. Each clustering starts with dissimilarity matrix $(D_k)_{j_1, j_2} = |\widehat \beta_{j_1,k} -\widehat \beta_{j_2,k} |,  0 \leq j_1,j_2\leq p_k.$ This matrix is consecutively updated as follows: a distance between two clusters $A$ and $B$ of levels of the $k$-th factor is defined as $\max_{a \in A, b \in B} |\hat \beta _{a,k} - \hat \beta _{b,k}|.$ Thus, each clustering begins with disjoint factor's levels and then two most {\it similar} clusters are merged. Finally, we obtain the factor with all levels merged.
Cutting heights from this clustering are contained in $h_k^T.$ Then we create vector $h$, which consists of elements of vector $(0,h_1^T, h_2^T, \ldots, h_{\hat S}^T)^T$ sorted increasingly. 
Now we  construct family $\mathcal{M}=\{\J _0=\hat S, M_1, M_2, \ldots , \{\emptyset\}\},$ where $\J _{j+1}$ is $\J _j$ with one additional merging of appropriate clusters corresponding to the $(j+1)$-th element in $h,$\\
{\bf (2b) Generalized Information Criterion   }
$$\hat {\J} _{PDMR}= \arg \min_{\J \in \mathcal{M}} \quad \ell(\hat \beta_ \J) + \lambda^2/2|\J|,
$$
where $\hat \beta_ \J$ is a minimum loss estimator over $\mathbb{R}^p$ with constraints determined by model $\J$ and $|\J|$ equals to $p$ minus a number of those constraints. Technical details of this constrained minimization is given in supplementary materials. We also show there that it can be  considered as an unconstrained minimization over a smaller space. \\
{\bf (2c) Estimation of parameters in model $\hat {\J} _{PDMR}$}
$$\hat{\beta}_{PDMR} = \arg \min_{\beta_{\hat \J _{PDMR}}} \ell (\beta_{\hat \J _{PDMR}})$$

\section{Statistical properties of PDMR}
\label{sec:properties}

We consider the PDMR algorithm with arbitrary diagonal matrices $W_k.$ The default setting from the previous section will be justified in Proposition \ref{opt_w}.

First, we generalize a characteristic of linear models with continuous predictors, which quantifies the degree of separation
between  model $M_{\bo}$ and other models \citep{YeZhang10}.
Let $S=\{1 \leq k \leq r: \bo _k \neq 0\}$  and $\bar S = \{1, \ldots, r\} \setminus S$. 
Notice that $S$ need not coincide with $M_{\bo}.$
For $ a \in (0,1)$ and a diagonal matrix $W$ we define a cone 
\begin{eqnarray}
 \label{cone}
{\cal C}_{a,W}=\{v \in \mathbb{R}^p:  
\sum_{k \in \bar S} ||W_k v_k||  \leq  
\sum_{k \in S}  ||W_k v_k|| +  a |W v|_1\}.
\end{eqnarray} 
A Cone Invertibility Factor (CIF)  is defined as
 \begin{equation}
 \label{CIF}
 \zeta_{a,W}=\inf_{0 \neq \nu\in {\cal C}_{a,W}}\frac {|W^{-1}X^TXv|_\infty}
{|\nu|_\infty}\:.                                         
 \end{equation}
In the case that matrix $X^TX$ is orthogonal one can easily find a lower bound on \eqref{CIF}. For instance, for the default choice of weights $W_k$ (i.e. $(W_k)_{jj}=||x_{j,k}||$) we have $\zeta_{a,W} \geq x_m$ for all $a \in (0,1),$ 
where we recall that $x_m=\min_{j,k} \; ||x_{j,k}||$ is the square root of the minimal number of observations per level.

In the case $n>p$ one usually uses the minimal eigenvalue of the matrix $X ^TX$ to express the strength of correlations between predictors. Obviously, in the high-dimensional scenario this value is zero. 
Therefore, CIF can be viewed as a useful analog of the minimal eigenvalue for the case $p>n.$
In comparison to more popular restricted eigenvalues \citep{BickelEtAl09} or compatibility constants \citep{GeerBuhlmann09}, CIF enables sharper 
$\ell_\infty$ estimation error bounds    \citep{YeZhang10, HuangZhang12, ZhangZhang12}. We explain  precisely this fact in supplementary materials. 
Finally, if all predictors are continuous, then \eqref{cone} and \eqref{CIF} are the same as the cone and CIF in \citet{YeZhang10}.

\subsection{Estimation consistency of Group Lasso}
\label{subsec:groupL}

Now we establish an upper bound on an estimation
error of the Group Lasso, which can be applied to the high-dimensional scenario  $p>>n.$ Similar results can be found in the literature, for instance in \citet[Theorem 4.5]{NardiRinaldo2008}, \citet[Theorem 2.2]{WeiHuang2010}, \citet[Theorem 5.1]{Lounici_etal2011}, \citet[Theorem 8.1]{BuhlmannGeer11} or \citet[Theorem III.6]{Blazere2014}. The main difference between those results and ours is that we measure the estimation error in the $l_\infty$ norm, which is all we need in partition selection, while in those papers there is a mixture of $l_2$ and $l_1$ (or $l_\infty$) norms.
Thus,  relying on those papers we would need  more restrictive assumptions in our results. It is especially true for an orthogonal design. At the end of this subsection  we compare our results to \citet[Theorem 5.1]{Lounici_etal2011}.

\begin{lemma}
\label{lem1}
Suppose that assumptions \eqref{inverseLink}, \eqref{subgauss} are satisfied and  $a \in (0,1).$  
Then
 \begin{equation*}
 \mathbb{P_{\bo}} \left(|\hat{\beta}-\mathring{\beta}|_\infty > (1+a)\lambda\zeta_{a,W}^{-1} \right) \leq 2p\exp\bigg(-\frac{a^2\lambda^2}{2\sigma^2 x^2_W}
 \bigg).
 \end{equation*}
Thus, if $\lambda ^2= 2 a^{-2}\sigma^2x^2_W \log(2p/\alpha)$ for some $\alpha \in (0,1),$ then  with probability at least $1-\alpha$
\begin{equation}
\label{bound_gl}
|\hat{\beta}-\mathring{\beta}|^2_\infty/\sigma^2 \leq 2(1+a)^2a^{-2} x_W^2 \zeta_{a,W}^{-2} \log(2p/\alpha). 
\end{equation}
\end{lemma}

The upper bound on the estimation error in Lemma \ref{lem1} depends on the choice of weight matrix $W.$
 So, to find optimal weights we should minimize $ x^2_W \zeta_{a,W}^{-2}.$ Solving this problem in the general case is difficult, so we restrict to the simplified version of the problem in the next result.
\begin{proposition}
\label{opt_w}
If $X^TX$ is orthogonal and weights are of the form $w_{j,k}=||x_{j,k}||^q$ for  $q \in \mathbb{R}.$ Then for each $a \in (0,1)$ we have
$
x^2_W \zeta_{a,W}^{-2}\leq x^{-2}_m (x_M/x_m)^{\max(0,|2q-3|-1)}=:f(q)$ and $\arg \min_q f(q)=[1,2].
$
\end{proposition}

Thus, for an orthogonal design with the optimal weights (i.e. $q \in [1,2]$) the upper bound in \eqref{bound_gl} behaves like $x^{-2}_m \log p.$ Consider a {\it balanced design}, i.e. there are $n/p_k$ observations on every level of $k$-factor. Then $x_m^{-2}=\max_k p_k/n$ and the upper bound on the estimation error of Group Lasso behaves like $ \sqrt{\max_k p_k \, \log p/n}.$ 
The assumption that a design is orthogonal is quite restrictive. The much more common case, especially for $p>>n,$ is an {\it almost orthogonal} design, i.e. ${x_{j_1,k_1}}^Tx_{j_2,k_2}=o(x_m^2)$ for $(j_1,k_1) \neq (j_2,k_2).$ In such a case weights 
$w_{j,k}=||x_{j,k}||^q$ for  $q \in [1,2]$ can be treated as {\it almost optimal}. 

In the original paper on the Group Lasso \citep{groupLasso} two choices of weights are proposed. The first one, called ``obvious'', gives a penalty 
of the form $\lambda \sum_k ||\beta_k||.$ In the second one, called ``preferred'' they have a penalty  $\lambda \sum_k \sqrt{p_k} ||\beta_k||.$ The latter choice is more widely used in the literature \citep{NardiRinaldo2008, WeiHuang2010, BuhlmannGeer11}.
Now we compare these choices of weights to those obtained in Proposition~\ref{opt_w}. Notice that columns of $X$ are normalized in \citet{groupLasso}, which is not done in our paper. So, we start with writing their penalty in our setting.
We do it under a balanced design (it is defined in the previous paragraph). Their first choice gives a penalty $ \lambda \sqrt{n} \sum_k p_k^{-1/2} ||\beta_k||,$ while the second one gives $ \lambda \sqrt{n} \sum_k  ||\beta_k||.$ On the other hand, by Proposition~\ref{opt_w} for $q=1$ we obtain
$ \lambda \sqrt{n} \sum_k p_k^{-1/2} ||\beta_k||,$ while for $q=2$ we have $ \lambda n \sum_k p_k^{-1} ||\beta_k||.$
Therefore, our optimal choice for $q=1$ coincides with the ``obvious'' choice in \citet{groupLasso}. However, the ``preferred'' choice in \citet{groupLasso} leads to sub-optimal results. Obviously, 
Proposition~\ref{opt_w} deals with an orthogonal design, so our result is rather a starting point of the thorough analysis on weights optimality. 

 Finally, 
notice that for an orthogonal and balanced design \citet[Theorem 5.1]{Lounici_etal2011}   bound  the estimation error of Group Lasso by $ x_m^{-2} (\max_k p_k + \log r), $ which is greater than $x^{-2}_m \log p$ in Lemma \ref{lem1}.

\subsection{Partition selection of PDMR}
\label{subsec:properties}

In this section we state the main theoretical result concerning our algorithm. First, we need to define the Kullback-Leibler (K-L) distance between true model $\T$ and its submodels. The precise definition of a submodel and its cardinality is given in supplementary materials. Roughly speaking, model $M$ is a submodel of $\T$ ($M \subsetneq \T $), if $M$ contains at least one additional merging of levels comparing to $\T .$  

Let $\J$ be a submodel of $\T$  and $k=|\T | - |\J |.$
Denote 
$$
\delta_k= ||X\bo - X_{\J} \beta^* _{\J}||^2,
$$
where $\beta^*_M =\arg \min _{\beta_M} ||X \bo - X_{\J} \beta_{\J}||^2.$
A scaled K-L distance between $\T$ and its submodels $\J$  is
 \begin{equation}
 \label{delta}
\delta_{\T}=\min_{k=1,\ldots,|\T|-1} \; \min_{\J: {\J}\subsetneq \T, |\T|-|\J|=k} \quad \frac{\delta _k}{k }\:.
 \end{equation}
Different variants of the K-L distance have been used in the consistency analysis of selection algorithms 
\citep[Section 3.1]{PokarowskiMielniczuk15}, but $\delta_{\T}$ defined in \eqref{delta} seems to lead to optimal results \citep[Theorem~1]{ShenEtAl13}. 

In the next theorem we establish properties of the PDMR algorithm in partition selection. We consider the default setting of weights in Group Lasso (i.e. $(W_k)_{jj} = || x_{j,k} ||$ ), so \eqref{CIF} simplifies to $\zeta_{a}.$

\begin{theorem}
\label{main_th}
Assume that there exists $0<a<1$  such that
\begin{equation}
\label{lambda_form} 
2 a^{-2}\sigma^2 \log (|\T |^2/(2\log 2))
 \leq \lambda^2 < \frac{\min(\Delta^2 \zeta_a^2, 4 \delta_{\T})}{16(1+a)^{2}}.
\end{equation}
 Then
 \begin{equation}
\label{prob}
 P( \hat M _{PDMR} \subsetneq \T)\leq (2p+|\T|^2)  \exp \left(-
\frac{a^2\lambda^2}{2 \sigma^2}
\right).
 \end{equation}
\end{theorem} 

Theorem \ref{main_th} states that the PDMR algorithm  is screening consistent, which means that with high probability it is able to reduce a model returned by the Group Lasso without losing any active variables. Indeed, taking $\lambda^2 = 2 a^{-2}\sigma^2 \log p (1+ o(1))$ we obtain that the right-hand side of 
\eqref{prob} is small, if the true model is sufficiently sparse, i.e. $|\T |^2 \leq p.$ Notice that it is the same choice of $\lambda$ as in the {\it Risk Inflation Criterion} \citep{FosterGeorge1994}. In the case that $|\T |^2 > p$ we can take slightly larger  $\lambda^2 = 4 a^{-2}\sigma^2 \log p (1+ o(1)).$

Obviously, the main result holds if the left-hand side in \eqref{lambda_form} is smaller than its right-hand side. However, this condition is not verifiable in practice, which is a common drawback of Lasso-based estimators and does not relate to the fact that predictors are categorical. Roughly speaking, condition \eqref{lambda_form} requires that true model $\T$ is sufficiently distinguishable from the others (i.e. $\Delta, \delta_{\T}$ are sufficiently large and $\zeta_a$ not too close to zero). 

The proof of Theorem \ref{main_th}  is given in supplementary materials. It can be sketched as follows: first we apply Lemma~\ref{lem1} to  Group Lasso. Then we establish that probability of choosing a submodel of $\T$ can be expressed as a Touchard polynomial and estimated using the recent combinatorial results from \citet{Ahle2022}. Obviously, we would like also to find an upper bound on $P(\T \subsetneq \hat M _{PDMR})$. However, our proof method fails in this case.  It relates to the fact that for categorical predictors a number of supermodels grows very quickly for $p>>n.$ However, we still believe that this problem might be solved in the future (possibly it has to wait for novel combinatorial results in the spirit of those from \citet{Ahle2022}).

\section{Experiments}
\label{sec:experiments}

In the theoretical analysis of Lasso-type estimators one usually considers only one value of tuning parameter $\lambda.$ We have also followed this way. However, the practical implementations can efficiently return   estimators for a data driven net of $\lambda$'s, as in the R~package {\tt glmnet}  \cite{FriedmanEtAl10}. Similarly, using a net of $\lambda$'s, the Group Lasso and the Group MCP algorithms have been implemented in the R~package {\tt grpreg} \citep{BrehenyHuang15}. In the paper we also propose a net modification of the PDMR algorithm based on the DMR implementation for high-dimensional data from the R~package {\tt DMRnet} \citep{DMRnet}. Namely, we implemented the following scheme:\\
1. For $\lambda$ belonging to the grid:
 (i) calculate the Group Lasso estimator $\hat \beta(\lambda)$,
(ii) perform complete linkage for each factor and  get nested family of models $M_1(\lambda)\subset M_2(\lambda)\subset \ldots$\\
2. For a fixed model dimension $c$, select a model $ M_c$ from  family  $(M_c(\lambda))_\lambda$, which has the minimal prediction loss.\\
3. Select a final model from sequence $(M_c)_c$ using  the Risk Inflation Criterion (RIC), see \citet{FosterGeorge1994}, i.e. using tuning parameter $2 \sigma^2 \log p,$ which is the same as suggested in Theorem \ref{main_th}.

Notice that PDMR uses the Group Lasso estimator $\hat \beta$ to compute dissimilarity measure, but DMR refits $\hat \beta$ with maximum likelihood and computes dissimilarity as the likelihood ratio statistics.

In experiments, in addition to PDMR and DMR, we evaluate also the following methods: \\
(i) Group MCP (\texttt{cv.grpreg} from the \texttt{R} package \texttt{grpreg} with \texttt{penalty="grMCP"} and with \texttt{gamma} set as the default for continuous response),\\
(ii) SCOPE from the \texttt{R} package \texttt{CatReg} \citep{stokell2021}. Tuning parameter \texttt{gamma} is chosen as 32, which is suggested in that paper for high-dimensional data.

\subsection{Simulation study}
\label{sec_sim}

This section contains the experiments with simulated high-dimensional linear models. In analyzed scenarios, design matrices $X$ and parameter vectors $\mathring{\beta}$ are the same as in \citet{stokell2021}, but additionally we systematically change the signal to noise ratio (SNR). For high-dimensional models,  consistency of partition selection is a very excessive requirement, so in the simulations we compare the relative prediction error PDMR and competitors with respect to the oracle depending on the difficulty of the prediction measured with  SNR.

In the training data we have $n=500$ observations. Every vector of predictors consists of $r=100$ factors, each on $24$ levels. Thus, after deleting 99 refence levels we obtain $p=2301.$ 
 A design matrix $X$ is generated as in \citet{stokell2021}, namely:  
first, we draw matrix $Z,$ whose rows $z_{i.},i=1,\ldots,500$ are independent $100$-dimensional vectors having  normal distribution $N(0,\Sigma).$ 
 The off-diagonal elements of $\Sigma$ are chosen such that correlation between $\Phi(z_{ij})$ and
$\Phi(z_{ik})$ equals $\rho$ for $j \neq k,$ where $\Phi$ is a cdf of the standard normal distribution.   In the paper we consider $\rho=0$ or $\rho=0.5.$ Then we set $x_{ij}=\lceil 24 \Phi(z_{ij})\rceil .$ 
 Finally, $X$ is recoded into dummy variables and reference levels of each factors, except the first one, are deleted.   

 Errors $\varepsilon_i$ are independently distributed from $N(0,\sigma^2),$ where $\sigma$ is chosen in such a way to realize distinct SNR. The performance of estimators is measured using root-mean-square errors (RMSE), which are calculated using the test data consisting of $10^5$ observations. To work on the universal scale  we divide RMSE of procedures by RMSE of the oracle, which knows the true model in advance. Final results are averages over 200 draws of training and test data. We consider the following six models, 
which are the same as in \citet[Section 6.1.2]{stokell2021}. But we renumber them with respect to true model dimensions (MD), i.e. a number of distinct levels among factors (recall that we do not count reference levels of factors, except the first one):\\
 {\bf Setting 1}: $\mathring{\beta} _k = (\underbrace{0,\ldots,0}_{7 \; \rm{times}}, \underbrace{2,\ldots,2}_{8 \; \rm{times}}, \underbrace{4,\ldots,4}_{8 \; \rm{times}})$ for $k=2,3,$ 
$\mathring{\beta} _k = (\underbrace{0,\ldots,0}_{15 \; \rm{times}},  \underbrace{5,\ldots,5}_{8 \; \rm{times}})$ for $k=4,5,6,$ $\mathring{\beta}_1=(0,\mathring{\beta}_2),$ and $\mathring{\beta} _k =0$ otherwise, $MD=10,$ \\
 {\bf Setting 2}: $\mathring{\beta} _k = (\underbrace{0,\ldots,0}_{7 \; \rm{times}}, \underbrace{2,\ldots,2}_{8 \; \rm{times}}, \underbrace{4,\ldots,4}_{8 \; \rm{times}})$ for $k=2,3,$ 
$\mathring{\beta} _k = (\underbrace{0,\ldots,0}_{9 \; \rm{times}}, \underbrace{2,\ldots,2}_{4 \; \rm{times}}, \underbrace{4,\ldots,4}_{10 \; \rm{times}})$ for $k=4,5,6,$ 
$\mathring{\beta}_1=(0,\mathring{\beta}_2),$ and  $\mathring{\beta} _k =0$ otherwise, $MD=13,$ \\
 {\bf Setting 3}: $\mathring{\beta} _k = (\underbrace{0,\ldots,0}_{5 \; \rm{times}}, \underbrace{2,\ldots,2}_{6 \; \rm{times}}, \underbrace{4,\ldots,4}_{6 \; \rm{times}}, \underbrace{6,\ldots,6}_{6 \; \rm{times}})$ for $k=2,...,5,$ 
 $\mathring{\beta}_1=(0,\mathring{\beta}_2),$ and $\mathring{\beta} _k =0$ otherwise, $MD=16,$\\
 {\bf Setting 4}: $\mathring{\beta} _k = (\underbrace{0,\ldots,0}_{4 \; \rm{times}}, \underbrace{1,\ldots,1}_{5 \; \rm{times}}, \underbrace{2,\ldots,2}_{4 \; \rm{times}}, \underbrace{3,\ldots,3}_{5 \; \rm{times}},\underbrace{4,\ldots,4}_{5 \; \rm{times}})$ for $k=2,...,5,$ 
 $\mathring{\beta}_1=(0,\mathring{\beta}_2),$ and $\mathring{\beta} _k =0$ otherwise,  $MD=21,$ \\
 {\bf Setting 5}: $\mathring{\beta} _k = (\underbrace{0,\ldots,0}_{3 \; \rm{times}}, \underbrace{2,\ldots,2}_{12 \; \rm{times}}, \underbrace{4,\ldots,4}_{8 \; \rm{times}})$ for $k=2,\ldots,10,$ 
$\mathring{\beta}_1=(0,\mathring{\beta}_2),$ and $\mathring{\beta} _k =0$ otherwise,  $MD=21,$  \\
 {\bf Setting 6}: $\mathring{\beta} _k = (\underbrace{0,\ldots,0}_{15 \; \rm{times}}, \underbrace{5,\ldots,5}_{8 \; \rm{times}})$ for $k=2,...,25,$ 
 $\mathring{\beta}_1=(0,\mathring{\beta}_2),$ and $\mathring{\beta} _k =0$ otherwise,  $MD=26.$ \\

In \citet[Section 6.1.2]{stokell2021} the above Settings 1-6 are numbered as 3, 1, 8, 4, 7 and 5, respectively. In each  model we consider $\rho=0$ or $\rho=0.5$ and distinct SNR, in particular
Settings 2 and 6 from that paper are also studied.

\begin{figure}[!htbp]
\centering
\includegraphics[width=1\columnwidth]{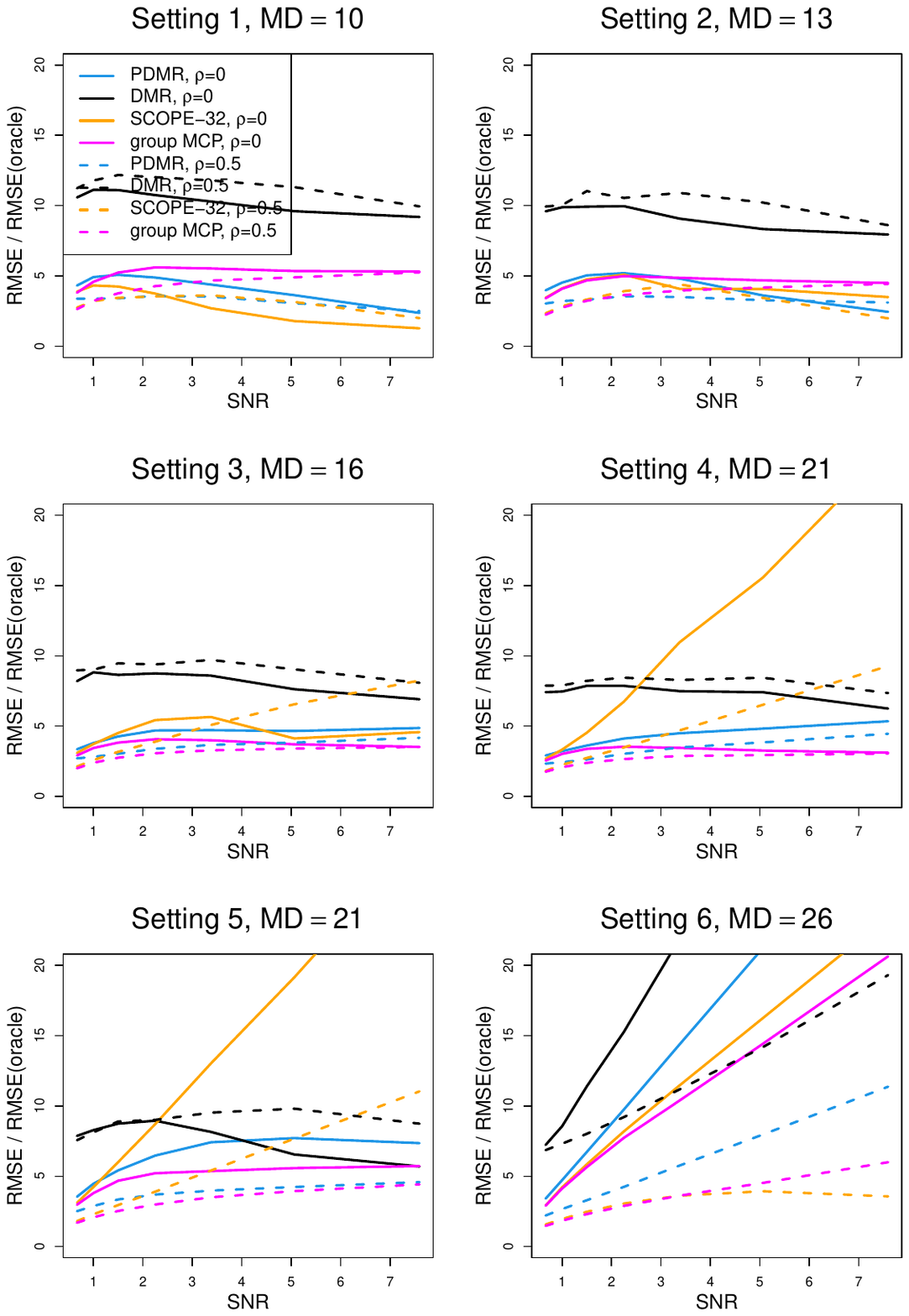} 
\caption{Relative prediction errors of considered methods. }
\label{fig2}
\end{figure}

The results of experiments are presented in Figure \ref{fig2}. For Setting 1 and~2 the relative prediction error of the algorithms that perform partition selection decreases as  SNR increases.  
The models induced by Settings 3, 4 and 5 are larger than the ones in Settings 1 and 2. It can be seen that in these tasks PDMR behaves better or significantly better than SCOPE and DMR. The predictive accuracy of PDMR is comparable or slighlty worse than MCP. However, MCP does not return sparse models. So, 
a price that MCP pays for good prediction is the difficulty in interpretation. 
Note that the most difficult Setting 6 does not satisfy the assumptions required for the Group Lasso operation. Indeed, the true factors have more parameters in total than the number of observations ($ n = 500$).

 \begin{figure}[!htbp]
\centering
\includegraphics[width=1\columnwidth]{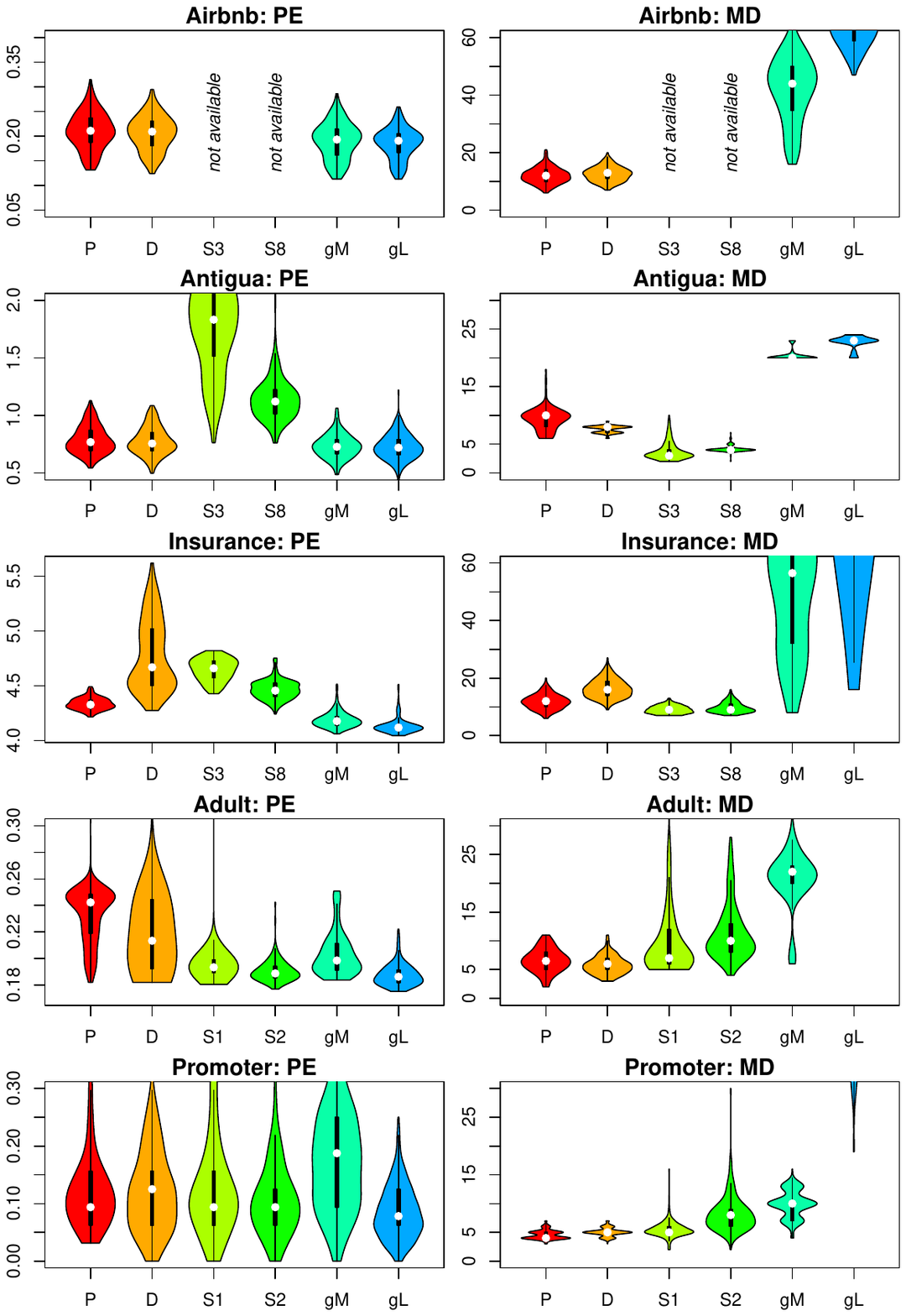} 
\caption{Results of experiments on real data. }
\label{fig3}
\end{figure}

\subsection{Real data study}
\label{sec:real}

We investigate five real data sets: first two with binary responses and the next three with continuous responses: \\
- the Adult data set~\citep{Kohavi} contains data from the 1994 US census. It contains 32,561 observations in a file \texttt{adult.data} and 16,281 observations in a file \texttt{adult.test}. The response  represents whether the individual's income is higher than 50,000 USD per year or not. We preprocessed the data as in~\citet{stokell2021}, i.e. we combined  two files together, removed 4 variables representing either irrelevant (\emph{fnlwgt}) or redundant (\emph{education-num}) features or with values for the most part equal to zero (\emph{capital-gain} and \emph{capital-loss}) and then removed the observations with missing values. Preprocessing resulted in 45,222 observations with 2 continuous and 8 categorical variables with $p=93$,\\
- the Promoter data set~\citep{Harley, Towell} contains E. Coli genetic sequences of length 57. The response  represents whether the region represents a gene promoter. We removed the \emph{name} variable and further worked with a data set consisting of 106 observations with 57 categorical variables, each with 4 levels representing 4 nucleotides, thus with $p=172$.
Both data sets are available at the UCI Machine Learning Repository~\cite{Dua_Graff},\\
- the Airbnb data set reports rental price for a number of hosts offering rental in an Airbnb service and is available from  {\it insideairbnb.com.} The host is characterised by a number of features like the neighbourhood, number of rooms, is it kids friendly, the length of rental history, reviews etc. We follow~\citet{simchoni_rosset21} in using the same dated version of the data set and in preprocessing the data as in~\citet{Kalehbasti2021}, i.e. we compute numeric sentiment for reviews or otherwise transform features into numbers and normalize them (including the log transformation of the rental price), with the following exceptions: (1) we retain the \emph{host\_id}, \emph{street} and \emph{neighbourhood} columns as categorical variables, with 39393, 311 and 204 levels, respectively and (2) we skip the feature selection step, since ability of considered methods to screen predictors is one of the things we want to test in this paper. This preprocessing procedure resulted in 49,976 observations with 767 variables (out of which 3 are categorical) with $p=40670$,\\
- the Insurance data set~\citep{Kaggle} contains data describing attributes of life insurance applicants. The response is an 8-level ordinal variable measuring insurance risk of the applicant, which we treat as a continuous response. We preprocessed the data as in~\citet{stokell2021}, i.e. we removed the irrelevant \emph{id} variable and 13 variables with missing values. Preprocessing resulted in 59,381 observations with 5 continuous and 108 categorical variables with $p=823$,\\
- the Antigua data set~\citep{Andrews} contains data concerning maize
fertilizer experiments on the Island of Antigua and is available at the \texttt{R} package \texttt{DAAG}~\cite{Maindonald}. The response measures harvest. We removed the irrelevant \emph{id} variable and one observation with a clearly outlying value of \emph{ears} variable and further worked with a data set consisting of 287 observations with 2 continuous and 3 categorical variables with $p=24$.

To study data with binary response variables we extend the PDMR algorithm from Section \ref{sec:model_alg} to logistic regression. From the practical point of view this generalization is relatively easy. First, we apply the Grooup Lasso for logistic regression, i.e. $\ell (\beta)=\sum_{i=1}^n [\log(1+\exp (x_{i.}^T \beta)) - y_ix_{i.}^T \beta ].$ A similar modification should  be done when applying the information criterion step. We have done it using  the R~package {\tt DMRnet}.

For a given data set and for each prediction method we performed 200 iterations of the following procedure:\\
- a sample of $m$ percent of a total number of observations was randomly selected from the data set. That sample was treated as a train set, \\
  - if there were any constant variables in the train set, those variables were removed from the train set,\\
    - the remaining $100-m$ percent of observations in the data set was treated as a test set for this iteration, with the following modifications: (1) we removed the same variables (columns) that had been removed from the train set, (2) we also removed from the test set the observations (rows) with levels not present in the train set,\\
    - the evaluated method was trained on a train set and the resulting model was used to obtain a prediction on a test set. This model and the prediction was then evaluated with the evaluation metrics. If either a training or a test phase couldn't be completed because of reported errors, that iteration was ignored and restarted with a new training sample.

The results are given in Figure \ref{fig3}:
PE - predition error, MD - model dimension, P - PDMR, D - DMR, S1, S2, S3, S8 - versions of SCOPE, gM - group MCP, gL  - group Lasso.
We do not show results for SCOPE for Airbnb data set because we were not able to complete the SCOPE computations with this dataset without errors for any \texttt{gamma} value considered (this is a continous response dataset, but nevertheless we tried 4 values: 8, 32, 100 and 250 for \texttt{gamma}, all with no success).

The algorithms for partition selection: PDMR, DMR and SCOPE have predictive accuracy slightly worse than the Group Lasso and Group MCP, but their outputs are significantly more sparse and have only a few parameters. Therefore, their results are much more interpretable. In general, the performance of PDMR and DMR is similar. For data sets with continuous response the SCOPE error is slightly larger than the PDMR error, but its model dimension is smaller. For Adult data, the opposite is true. 
This difference can probably be reduced by mixing the PE and MD criteria for final selection.

\section{Conclusions}

We present PDMR, that is an improvement of DMR  in terms of performance and simplification for theoretical analysis. In simulation experiments PDMR gives a prediction error significantly lower than DMR and 
 lower than SCOPE, but in experiments with real data, all three algorithms perform similarly. SCOPE is several dozen times slower than PDMR and DMR.

For high-dimensional scenario there are no theoretical results regarding the consistency of the DMR selection, while for SCOPE, a weaker property than selection consistency was proved in \citet{stokell2021}. Our result is also weaker than the selection consistency, but it  seems easier to interpretation.

The algorithms for partition selection: PDMR, DMR and SCOPE have predictive accuracy slightly worse than the Lasso, Group Lasso and Group MCP, but their outputs are significantly more sparse and have only a few parameters. Therefore, their results are much more interpretable.

As a by-product we obtain optimal weights for the Group Lasso, which are different from those recommended by the authors of this method. Possibly, the new weights can improve asymptotics of the Group Lasso in the general scenario (not necessarily orthogonal) and its practical performance as well.

There are two theoretical extensions of our paper. The first one is to prove an analogous bound to that in Theorem 1 but concerning supermodels of $ \T$  The second one is to
generalize this theorem, for instance to GLMs.

\appendix

\section*{Supplementary materials}

\section{Missing proofs}

\subsection{Proof of Lemma \ref{lem1}}

For $k=1,\ldots, r$ using KKT for  the Group Lasso estimator $\hat{\beta},$ we have that $W_k^{-1} \ml _k (\hat{\beta})= 
 -\lambda W_k \hbeta _{k}/||W_k \hbeta _k|| $ for $\hbeta _k \neq 0$ and $||W_k^{-1} \ml _k (\hat{\beta})|| \leq
 \lambda $ for $\hbeta _k = 0.$ Therefore, we obtain that $|W^{-1} \ml  (\hat{\beta})|_\infty= \max _{k} |W_k^{-1} \ml _k (\hat{\beta})|_\infty \leq \lambda.$

Recall that $\ml (\bo)=-X^T\varepsilon$ and
suppose that we are on  event $\mathcal{A}=\{|W^{-1}\ml (\bo)|_\infty\leq a\lambda\}.$
First,  we  prove that $v:=\hat \beta - \bo \in {\cal C}_{a,W}.$ Using  differentiability of $\ell$ and Taylor's expanssion we have $ v ^T \left[  \dot \ell (\hbeta) - \dot \ell  (\bo)\right]= v^T \nabla^2 \ell (\tilde \beta) v$ for some $\tilde \beta$ between $\hbeta$ and $\bo.$ Obviously, this expression is nonnegative, because $\ell$ is convex.
Moreover, $v _k =\hbeta _k$ for $k \in \bar S,$ so we also obtain
\begin{eqnarray}
\label{form1}
 v ^T \left[  \dot \ell (\hbeta) - \dot \ell  (\bo)\right]=
\sum_{k=1}^r v _k ^T \dot \ell _k (\hbeta) - \sum_{k=1}^r v _k ^T \dot \ell _k (\bo)  
= \sum_{k \in \bar S} \hbeta _k ^T \dot \ell _k (\hbeta) + \sum_{k \in S} v _k ^T \dot \ell _k (\hbeta) 
 - \sum_{k=1}^r v _k ^T \dot \ell _k (\bo).
\end{eqnarray}
Consider the first term in \eqref{form1}. Using KKT, it equals
\begin{eqnarray*}
 \sum_{k \in \bar S,\hbeta _{k} \neq 0} 
 \hbeta _{k} ^T \dot \ell _{k} (\hbeta)
= -\lambda \sum_{k \in \bar S,\hbeta _{k} \neq 0}  \;||W_k\hbeta _k||  =
-\lambda \sum_{k \in \bar S}  \;||W_kv _k||.
\end{eqnarray*}
Similarly, we bound the second term in \eqref{form1} by
\begin{eqnarray*}
 \sum_{k \in  S} 
 ||W_k v _k||\; || W_k ^{-1}\dot \ell _{k} (\hbeta)||  \leq \lambda \sum_{k \in  S}||W_k v _k||.
\end{eqnarray*}
The last term in \eqref{form1} can be bounded using the fact that we are on  event $\mathcal{A}$ 
$$
\sum_{k=1}^r |W_k v _k|_1 |W_k^{-1} \dot \ell _k (\bo)|_\infty 
\leq a\lambda \sum_{k =1}^r |W_k v _k|_1.
$$
Joining the above facts we get that $v \in  {\cal C}_{a,W}.$
Therefore, from the definition \eqref{CIF} we have
\begin{eqnarray*}
 \zeta_{a,W}|\hbeta - \bo|_\infty \leq  
\max\limits_{1\leq k \leq r} |W_k^{-1}\dot \ell _k (\hbeta)
-W_k ^{-1} \dot \ell _k (\mathring{\beta})|_\infty  
\leq \max_{1\leq k \leq r} 
 |W_k ^{-1}\dot \ell _k (\hbeta)|_\infty
+\max_{1\leq k \leq r} 
|W_k^{-1} \dot \ell _k (\mathring{\beta})|_\infty .
\end{eqnarray*}
Using again KKT and the fact, that we are on $\mA ,$ we get $|\hat{\beta}-\mathring{\beta}|_\infty \leq (1+a)\lambda\zeta_{a,W}^{-1}.$ Now we calculate probability of event~$\mA.$
To do it, we use the following exponential inequality for independent subgaussian variables $\varepsilon_i, i=1,\ldots ,n$: for each $b>0$ and $v\in \mathbb{R}^n$ we have $P(\varepsilon^Tv/||v||>b) \leq \exp\left(-b^2/(2\sigma^2)\right).$ 
Using union bounds and the definition of $x_W,$ we obtain
\begin{eqnarray*}
P_{\bo}(\mA ^c) \leq
 \sum_{k,j} P\left(|x_{j,k}^T \varepsilon|/w_{j,k} > a\lambda \right) 
\leq 2 \sum_{j,k} \exp \left(
-\frac{a^2 \lambda^2 w_{j,k}^2}{2 \sigma^2 ||x_{j,k}||^2} 
\right) 
 \leq
2p  \exp \left(
-\frac{a^2 \lambda^2 }{2 \sigma^2 x_W^2} \right),
\end{eqnarray*}
where we consider $k \in \{1,\ldots, r\}$ and for $k=2,\ldots,r$ we have  $j \in \{1,\ldots,p_k\}, $ while for $k=1$ we have $j \in \{0,\ldots,p_1\}. $

The proof of the second claim is straightforward.

\subsection{Proof of Proposition \ref{opt_w}}

For a linear model with an orthogonal design  we have $|W^{-1} X^TX v|_\infty/|v|_\infty \geq \min_{j,k}||x_{j,k}||^{2-q}$ for all $v \in \mathbb{R}^p.$ So, we can easily bound from above $
\zeta_{a,W}^{-2}
$ by $x_m^{2q-4},$ when $q\leq 2$ and $x_M^{2q-4},$ when $q> 2.$  The rest of the proof follows from the fact that $x^2_W$ equals $x_M^{2-2q},$ when $q\leq 1$ and $x_m^{2-2q},$ when $q> 1.$

\subsection{Proof of Theorem \ref{main_th}}

We will  establish two inequalities 
 \begin{equation}
\label{eq_screen}P(\T \notin \mathcal{M}) \leq 2p \exp \left(-
\frac{a^2\lambda^2}{2 \sigma^2}
\right)
\end{equation}
and
\begin{equation}
\label{subs}
 P(\T \in \mathcal{M}, \hat M _{PDMR} \subsetneq \T)\leq (2 \log 2)^{-1} |\T| ^2\exp\bigg(-\frac{a^2\lambda^2}{2\sigma^2}\bigg).
 \end{equation}

We start with \eqref{eq_screen}. From Lemma \ref{lem1} we know that 
$$P(|\hat \beta - \bo| _\infty \leq  (1+a)\lambda \zeta_a^{-1}) \geq 1-2p \exp \left(-
\frac{a^2\lambda^2}{2 \sigma^2}
\right).
$$

Now we fix the $k$-th predictor and take indexes $j_1,j_2$ such that $\bo _{j_1,k}= \bo _{j_2,k},$ i.e. they correspond to the same cluster in $\T$. Let $R=(1+a)\lambda \zeta_a^{-1}.$ We obtain
\begin{equation}
|\hbeta _ {j_1,k} - \hbeta_{j_2,k}|\leq |\hbeta _ {j_1,k} - \bo _{j_1,k}| + |\hbeta _ {j_2,k} - \bo _{j_2,k}| 
\leq 2R.
\end{equation}
On the other hand, if $j_1,j_2$ are such that
$\bo _{j_1,k} \neq \bo _{j_2,k}$, then
\begin{eqnarray}
|\hbeta _ {j_1,k} - \hbeta_{j_2,k}|  
\geq |\bo _{j_1,k} -\bo _{j_2,k}| - |\hbeta _ {j_1,k} - \bo _{j_1,k}| - |\hbeta _ {j_2,k} - \bo _{j_2,k}|
\geq \Delta  -2R >2R,
\end{eqnarray}
because $\Delta > R$ by assumption \eqref{lambda_form}.
Therefore, there is a separation between entries of dissimilarity matrix $D_k$ in the clustering step of PDMR. Namely, entries corresponding to indexes from the same true cluster are smaller than those corresponding to distinct true clusters.
The first step of complete linkage clustering uses dissimilarity matrix $D_k,$ then in consecutive steps
this matrix is updated as follows: a distance between two clusters $A$ and $B$ is defined as $\max_{a \in A, b \in B} |\hat \beta _{a,k} - \hat \beta _{b,k}|.$ Therefore,  in some step of  clustering we obtain true partitioning of levels of the $k$-th factor and all cutting heights in $h_k$ to that step are not larger than $2R$, while subsequent coefficients of $h_k$ are larger than $2R.$ Clearly, threshold $2R$ does not depend on $k.$ Therefore, this separation property is also satisfied after taking all cutting heights together and sorting increasingly. Thus, the true model  is contained in  nested family $\mathcal{M}$ with high probability.

Now, we consider \eqref{subs}. Notice that
\begin{equation}
\label{ls_f1}
P(\T \in \mathcal{M}, \hat M_{PDMR} \subset \T) \leq P \left( \exists_{M: L_{\J} \subsetneq L_{\T}} \quad 
\ell(\hat\beta_M)
 +\lambda^2 |M|/2 < \ell(\hat \beta_{\T}) + \lambda^2 |\T|/2
\right)
\end{equation}
and we recall that $\hat \beta_M$ is  a minimum loss estimator over $\mathbb{R}^p$ with constraints determined by model $\J .$ Technical details of this constrained minimization is given in Section \ref{sec:models} of these supplementary materials.
Denote $k=|\T |-|M|.$ We can calculate that $\delta_k = ||(I-H_M)X \bo ||^2$ and
$$
\ell(\hat \beta_M) = \delta_k/2 + \varepsilon ^T(I-H_M)X \bo  + \varepsilon ^T (\mathbb{I} - H_M) \varepsilon/2 - y^Ty/2,
$$
in particular $\ell(\hat \beta_{\T})=\varepsilon ^T (\mathbb{I} - H_{\T}) \varepsilon/2 - y^Ty/2.$ Since $H_{\T} - H_M$ is a projection matrix, we have
$$
\ell(\hat \beta_M) - \ell(\hat \beta_{\T}) \geq \delta_k/2 + \varepsilon ^T(I-H_M)X \bo 
$$
and  we can bound the rhs of \eqref{ls_f1} by
$$
 P \left( \exists_{M: L_M \subsetneq L_{\T}} \quad 
-2 \varepsilon ^T(I-H_M)X \bo  \geq \delta_k - k\lambda^2 \right).
$$
Clearly, we have $\delta_k \geq k\delta_{\T},$ so above probability is bounded by 
\begin{equation}
\label{ls_f2}
P \left( \exists_{M: L_M \subsetneq L_{\T}} \quad 
\frac{- \varepsilon ^T(I-H_M)X \bo }{\sqrt{\delta_k}} \geq \sqrt{k\delta_{\T}}\left(1 - 
\frac{\lambda^2}{\delta_{\T}}\right)/2 \right).
\end{equation}
To estimate \eqref{ls_f2} we use union bounds with the exponential inequality for subgaussian random variables (see the proof of Lemma \ref{lem1}). Thus, we bound \eqref{ls_f2} from above by
\begin{equation}
\label{ls_f3}
 \sum_{k=1}^{|\T|-1} N_k
 \exp\left({-\frac{k \delta_{\T}}{8\sigma^2}\bigg(1-\frac{\lambda^2}{ \delta_{\T}}\bigg)^2}\right),
\end{equation}
where $N_k$ is a number of models $M$ such that $L_M \subsetneq L_{\T}$ and $|M|=|\T|-k.$ Notice that the value in \eqref{ls_f3} is the largest, if $\T$ consists of one factor on $|\T|$ levels, which we assume in the following. In this case $N_k = {|\T| \brace |\T|-k },$ where ${r \brace s}$ is a Stirling number of the second kind, i.e. a number of ways to partition a set of $r$ objects into $s$ non-empty subsets.

From the assumption $\lambda^2 \leq \delta_{\T}/(2+2a)^2$ we obtain 
\begin{equation}
\label{quadratic}
\lambda^2/\delta_{\T}\leq f_1(a), \quad\quad {\rm where}\quad f_1(a)=1 +2a^2-\sqrt{(1+2a^2)^2-1},
\end{equation}
which gives 
\begin{equation}
\label{ls_f4}
\frac{4a^2\lambda^2 }{\delta _{\T}}\leq \bigg(1-\frac{\lambda^2}{ \delta_{\T}}\bigg)^2.
\end{equation}
Therefore, we estimate \eqref{ls_f3} by    
\begin{equation}
\label{ls_f5}
 \sum_{k=1}^{|\T|-1} {|\T| \brace |\T|-k}
 \exp\left(-\frac{k a^2 \lambda ^2}{2\sigma^2}\right) = 
\exp\left(- |\T| \frac{a^2 \lambda ^2}{2\sigma^2}
\right)
\sum_{k=1}^{|\T|} {|\T| \brace k}
 \exp\left(\frac{k a^2 \lambda ^2}{2\sigma^2}\right) -1.
\end{equation}
The sum in \eqref{ls_f5} is called a Touchard polynomial. Its value is closely related to moments of Poisson random variables (see Lemma \ref{poiss} given below). Therefore,  \eqref{ls_f5} can be estimated by Lemma \ref{pois_bound} (given below) as
$$
\exp \left[|\T|^2\exp\left(-  \frac{a^2 \lambda ^2}{2\sigma^2}
\right)/2\right]-1.
$$

Using the inequality $\exp(c)-1 \leq \log (2)^{-1}c$ for $0\leq c \leq \log (2),$ we finish the proof.

\begin{lemma}[\cite{PeccatiTaqqu2011}, Proposition 3.3.2]
\label{poiss}
For every $n \geq 0$ and $x>0$, one has that
$\Ex [K(x)]^n = \sum_{k=1}^{n} {n \brace k} x^k,$
where $K(x)$ is a Poisson random
variable with parameter $x.$
\end{lemma}

\begin{lemma}[\cite{Ahle2022}, Theorem 1] 
\label{pois_bound}
Under assumptions of Lemma \ref{poiss} (given just above) we have
$$
\Ex [K(x)/x]^n \leq \left(\frac{n/x}{\log(1+n/x)}
\right)^n \leq \exp(n^2/(2x)).
$$ 
\end{lemma}

\section{Models and constrained minimization}
\label{sec:models}

For simplicity, we consider the case that all predictors are factors. But the extension to the general case is straighforward.

We recall that each model $M$ is defined as a sequence $M=(P_1,P_2,\ldots,P_r),$ where $P_k$ is some partition of a set of levels of the $k$-th factor, i.e. $\{0,1,\ldots, p_k\}.$
We will show that every model $M$ corresponds to linear space
\begin{equation}
\label{LJ}
L_ \J=\{\beta \in \mathbb{R}^p: A_{0,M} \beta=0\},
\end{equation}
where matrix $A_{0,M}$  is defined in the following subsection.

\subsection{Matrix $A_{0,M}$}

  Suppose that $P_k = C_{1,k} \cup C_{2,k} \cup C_{j_k,k},$ so $j_k$ is a number of clusters that set $\{0,1,\ldots, p_k\}$ is divided. In further considerations we fix the ordering between clusters. We also suppose that reference levels (i.e. zero levels) belong to $C_{1,k}$ for each $k.$  Let $s_{j,k}$ be the smallest element in $C_{j,k}.$ In particular, $s_{1,k}=0.$

Fix $\beta \in \mathbb{R}^p.$ We recall that $\beta=(\beta _1^T, \beta _2^T, \ldots, \beta _r^T
)^T,$ where $\beta _1 = (\beta _{0,1}, \beta _{1,1}, \ldots,
\beta _{p_1,1} )^T\in \mathbb{R}^{p_1+1}$ and  $\beta _k = (\beta _{1,k}, \beta _{2,k}, \beta _{3,k}, \ldots,
\beta _{p_k,k} )^T\in \mathbb{R}^{p_k}$ for  $k=2,\ldots, r.$ Now we change the ordering between coordinates of $\beta$ according to the ordering defined by model $M$ in the following way:
\begin{equation}
\label{bbeta}
\beta = (\underbrace{\beta_{s_{1,1},1},\beta_{s_{2,1},1},\ldots, \beta_{s_{j_1,1},1}}_{group \; 1}, \underbrace{\beta_{s_{2,2},2},\ldots, \beta_{s_{j_2,2},2}}_{group \;2}, \ldots,\underbrace{ \beta_{s_{2,r},2},\ldots, \beta_{s_{j_r,r},r}}_{group \; r}, \quad  {\rm remaining \; coefficients}).
\end{equation}
In other words, levels of the first factor are partitioned as $\{0,1,\ldots, p_1\}=C_{1,1} \cup C_{2,1} \cup C_{j_1,1}$ and the smallest numbers in these clusters are $s_{1,1}, s_{2,1}, \ldots, s_{j_1,1},$ respectively. So, {\it group 1} consists of the corresponding coefficients of $\beta_1 \in \mathbb{R}^{p_1+1}.$   Next, levels of the second factor are partitioned as $\{0,1,\ldots, p_2\}=C_{1,2} \cup C_{2,2} \cup C_{j_2,2}$ and the smallest numbers in these clusters are $s_{1,2}, s_{2,2}, \ldots, s_{j_2,2},$ respectively. So, {\it group 2} consists of the corresponding coefficients of $\beta_2 \in \mathbb{R}^{p_2}.$ In particular, {\it group 2} does not contain $\beta_{s_{1,2},2}$,  because $s_{1,2}$ corresponds to a cluster, which contains a reference level of the second factor and we do not include coefficients corresponding to reference levels in vector $\beta$ (cf. Section \ref{sec:model_alg} in the main part of the paper, the only exception is a reference level of the first factor). The same proceeding relates to the following factors. At the end, we write all coefficients, which were not used before. They are called {\it remaining  coefficients} in \eqref{bbeta}.

To make this new ordering more transparent we consider the example that we have two factors: the first one with 8 levels and the second one with 7 levels. So, $p_1=7,p_2=6$ and $p=14.$ Let $M=(P_1,P_2)$ be as follows: $P_1=\{0,2,6\} \cup \{3,4,5\}\cup\{1,7\},$
$P_2=\{0,4\} \cup \{1,2,6\}\cup\{3,5\}.$  Then $$\beta=(\beta_{0,1},\beta_{3,1},\beta_{1,1},\beta_{1,2}, \beta_{3,2},
\beta_{2,1},\beta_{6,1},\beta_{4,1},\beta_{5,1},\beta_{7,1},\beta_{4,2},\beta_{2,2},\beta_{6,2},\beta_{5,2}). $$

Let $m$ be a number of clusters indicated by model $M,$ which do  not contain reference levels plus one, i.e. $m=j_1+(j_2-1) + \ldots+ (j_k-1).$
Matrix $A_{0M}$ is a $(p-m)\times p$ matrix 
 of a form
$(B_M, \mathbb{I}_{p-m})$ for $(p-m)\times m$ matrix $B_M$ and identity matrix $ \mathbb{I}_{p-m},$ where matrix $B_M$ is constructed 
as follows: first, we define a connection between columns of $B_M$  and the first $m$ coordinates of \eqref{bbeta} as follows: the first column of $B_M$ corresponds to $\beta_{s_{1,1},1},$ the second one corresponds to $\beta_{s_{2,1},1}$ etc. Analogously, columns of 
$\mathbb{I}_{p-m}$ correspond to the last $p-m$ coordinates of \eqref{bbeta} (i.e. those called {\it remaining  coefficients}). Now we find any $1$ in matrix $\mathbb{I}_{p-m},$ say it is in column $t^*$ and row $t^*$ of matrix $\mathbb{I}_{p-m}.$  This column corresponds to some coordinate in \eqref{bbeta}, say $\beta_{j^*,k^*}.$ It means that this column corresponds to the $j^*$-th level of the $k^*$-th factor. Now we check to which cluster this level belongs to. Then we find the smallest element in this cluster, say $r^*$. If $r^* \neq 0$ or $r^*=0$ but $k^*=1$, then we take matrix $B_{M}$ and write $-1$ in its column corresponding to coordinate $\beta_{r^*,k^*}$ and the row $t^*$. The remaining entries in $B_{M}$ are filled in by zeroes.

In the above example we have $m=5$ and
\[
A_{0,M} = \begin{array}{ccccc|ccccccccc}
\beta_{0,1}& \beta_{3,1} & \beta_{1,1} & \beta_{1,2} & \beta_{3,2} & \beta_{2,1} & \beta_{6,1} & \beta_{4,1} & \beta_{5,1} & \beta_{7,1} & \beta_{4,2} & \beta_{2,2} & \beta_{6,2}& \beta_{5,2}\\ 
\left(\begin{array}{c}   -1 \\ -1 \\ 0 \\ 0 \\ 0 \\ 0 \\ 0 \\ 0 \\ 0\\ \end{array}\right.
&
\begin{array}{c} 0 \\ 0 \\ -1 \\ -1 \\ 0 \\ 0 \\ 0 \\ 0 \\ 0\\ \end{array}
&
\begin{array}{c} 0 \\ 0 \\ 0 \\ 0 \\ -1 \\ 0 \\ 0 \\ 0 \\ 0 \\\end{array} 
&
\begin{array}{c} 0 \\ 0 \\ 0 \\ 0 \\ 0 \\ 0 \\ -1 \\-1 \\ 0 \\ \end{array}
&
\begin{array}{c}0 \\ 0 \\ 0 \\ 0 \\ 0 \\ 0 \\ 0 \\0  \\ -1 \\\end{array} 
&
\begin{array}{c} 1 \\ 0 \\ 0 \\ 0   \\ 0 \\0 \\ 0 \\0 \\ 0 \\ \end{array} 
&
\begin{array}{c} 0 \\ 1 \\ 0 \\ 0 \\ 0 \\ 0 \\ 0 \\0 \\ 0 \\ \end{array}
&
\begin{array}{c} 0 \\ 0 \\1 \\ 0 \\ 0 \\ 0 \\ 0 \\ 0 \\ 0 \\ \end{array}
&
\begin{array}{c}  0 \\0\\0 \\ 1 \\ 0 \\ 0 \\ 0 \\ 0 \\ 0 \\ \end{array}
&
\begin{array}{c} 0 \\ 0 \\0 \\ 0 \\ 1 \\ 0 \\ 0 \\ 0 \\ 0 \\ \end{array}
&
\begin{array}{c} 0\\0 \\ 0 \\0  \\ 0 \\ 1 \\ 0 \\ 0 \\ 0 \\ \end{array}
&
\begin{array}{c} 0 \\ 0 \\0 \\ 0 \\ 0 \\ 0 \\ 1 \\ 0 \\ 0 \\ \end{array}
&
\begin{array}{c} 0 \\ 0 \\0 \\ 0 \\ 0 \\ 0 \\ 0 \\ 1 \\ 0 \\ \end{array}
&
\left.\begin{array}{c} 0 \\ 0 \\0 \\ 0 \\ 0 \\ 0 \\ 0 \\ 0 \\ 1 \\ \end{array}\right)
\end{array}.
\]
Therefore, space $L_M$ defined in \eqref{LJ} consists of those vectors $\beta,$ which determines the same partitions of factors' levels (and the same as given by $M$). These partitions are also coded by $A_{0,M}$.

\subsection{Constrained minimization}

In the paper we have many places where we consider {\it $\hat \beta_M,$ which is a minimum loss estimator over $\mathbb{R}^p$ with constraints determined by model $M$.} Now we can precisely define this minimization as $\arg \min _{\beta \in L_M} \ell (\beta).$ In this subsection we show that this constrained minimization can be replaced by the unconstrained one. 

Let $A_{1,M}=(\mathbb{I}_m, {\bf 0}_{m \times (p-m)})$ be  $(m \times p)$-complement of $A_{0,M}$ to invertible matrix $A_{M}$, that is:
$$
A_M=\left[\frac{A_{1,M}}{A_{0,M}}
\right] \quad {\rm and} \quad 
A_M^{-1}=\left[A_{M}^1 | A_{M}^0
\right]=\left[\substack{I_{m} \\ -B_{M}} | \substack{{\bf 0}_{m \times (p-m)} \\ \mathbb{I}_{p-m}}
\right],
$$
where $A_M^{-1}$ is calculated using the Schur complement.

Let $\beta_M $ be an arbitrary element of $L_M$  and $\xi _M = A_{1,M}\beta_M,$ then 
$\beta_M = A^1_M \xi_M.$
Indeed, we have
$$
\beta_M=A_M^{-1}A_M \beta_M = A_M^{-1} \left[\frac{A_{1,M} \beta_M}{A_{0,M}\beta_M}
\right] = \left[A_{M}^1 | A_{M}^0\right] \left[\frac{\xi_M}{{\bf 0}_{p-m}}
\right] =  A^1_M \xi_M.
$$
Therefore,   $L_M$ in \eqref{LJ} can be equivalently expressed as
$$
L_M=\{A^1_M \xi: \xi \in \mathbb{R}^m\}.
$$
Therefore, $L_M$ can be viewed as a linear space spanned by columns of $A^1_M.$ The dimension
of space $L_M$ is called a size of model $M$ and denoted by $|M|.$ Clearly, we have $|M| = m$, so $|M|$ is a number of different non-reference levels (again with an exception for the first factor) indicated by model $M.$

Fix model $M$ we change the ordering of columns in $X$ according to the ordering induced by $M,$ as in \eqref{bbeta}. Then matrix $Z_M=X A^1_M$ is simply matrix $X$ with appropriate columns deleted or added to each other according to partitions in model $M=(P_1,\ldots, P_r).$
We also have $X \beta_M=Z_M\xi _M.$ 
We assume that the considered  models are sufficiently sparse, which means that $r(Z_M)=|M| \leq \bar m,$ where $\bar m < \min (n,p)$ and $r(Z_M)$ is the rank of $Z_M.$ 

Therefore, constrained minimization $\arg \min _{\beta \in L_M} \ell (\beta)$ can be replaced by the unconstrained one as follows: we compute an ordinary least squares estimator with design matrix $Z_M,$ i.e.
$\hat \xi _M = (Z_M ^TZ_M)^{-1} Z_M^Ty.$ Then we calculate $ \hbeta _M = A^1_M \hat \xi _M.$

Finally, we also need the following notation of a projection matrix $H_M = Z_M (Z_M ^TZ_M)^{-1} Z^T_M.$ 

\subsection{Submodels}

Now we can easily define submodels, namely $M_1$ is a submodel of $M_2$, if $L_{M_1} \subset L_{M_2}.$ Roughly speaking, partitions induced by $M_1$ and $M_2$ are the same ($L_{M_1}=L_{M_2}$) or partitions induced by $M_1$ contains at least one additional merging of levels comparing to those induced by $M_2$ ($L_{M_1} \subsetneq L_{M_2}$).
Finally, a model determined by vector $\beta$ is such $M$ that $\beta \in L_{M}$ and $M$ has the smallest size among all models with this property. It is denoted $M_\beta .$

\section{Cone invertibility factor (CIF)}

Consider linear model \eqref{inverseLink} with numerical predictors only.
Let  $T$ be the set of indices corresponding to the support of the true vector $\bo$ and $T'=\{1,\ldots,p\} \setminus T.$ Let $\beta_T$ and $\beta_{T'}$ be the restrictions of the vector $\theta \in \mathbb{R}^p$ to the indices from $T$ and $T',$ respectively. Now, for $a \in (0,1)$ we consider cone 
$$\cone _a = \{ \theta \in \mathbb{R}^p : |\theta_{T'} |_1 \leq \frac{1+a}{1-a} |\theta_{T} |_1\}\;\;.
$$

In the case when $p>>n$ three different characteristics measuring the potential for consistent estimation of the model parameters have been introduced:\\
- the restricted eigenvalue \citep{BickelEtAl09}:\\
$$
RE_a = \inf_{0 \neq \theta \in \cone _a} \frac{\theta ^T X^TX\theta}{|\theta|^2_2}\;\;,
$$
- the compatibility factor \citep{BuhlmannGeer11}:
$$
K_a=\inf_{0 \neq \theta \in \cone _a } \frac{|T|\theta ^T X^TX\theta}{|\theta_T|_1^2}\;\;,
$$
- the cone invertibility factor (CIF, \citep{YeZhang10}): for $q\geq 1$
$$
\zeta_{a,q} = \inf_{0 \neq \theta \in \cone _a } \frac{|T|^{1/q}|X^TX\theta|_\infty}{|\theta|_q}\;\;.
$$
Relations between the above quantities are discussed, for instance, in \citet{GeerBuhlmann09, YeZhang10, Cox13}. Moreover, notice that $\zeta _{a,\infty}$ is the same as \eqref{CIF}, if we omit weight matrix $W.$

In this article we will use CIF, since this factor allows for a sharp formulation of convergency results for all  $l_q$ norms with $q\geq 1.$ Indeed, the following estimation bounds are established for Lasso with numerical predictors (see \citet[Section 3]{Cox13}): with probability close to one
\begin{eqnarray}
\label{cc}
|\hat \beta - \bo|_1 &\leq& \frac{2(1+a)|T| \lambda}{
(1-a)K_a} \quad =:R^1_a \\
\label{re}
|\hat \beta - \bo|_2 &\leq& \frac{(1+a)|T|^{1/2} \lambda}{
RE_a} \quad =: R^2_a  \\
\label{ccif}
|\hat \beta - \bo|_q &\leq& \frac{(1+a)|T|^{1/q} \lambda}{
\zeta_{a,q} } \quad =: R^3_{a,q}
\end{eqnarray}
Such estimation bounds are the main tool to prove selection consistency of modifications of Lasso such as
Thresholded Lasso, Adaptive Lasso or algorithms with nonconvex penalties (SCAD, MCP). Indeed, these inequalities are used to prove {\it separability} of Lasso, i.e. for each $j \in T$ and $k \notin T$ we have $|\hat \beta _j| \geq |\hat \beta_k|.$ To get it one has to assume additionally that $\bmin=\min_{j \in T} |\bo _j|$ is bounded from below by twice the right-hand side of \eqref{cc}, \eqref{re} or \eqref{ccif}.
The latter condition means that signal has to be large enough. Obviously, one wants this condition to be as weak as possible. Below we show that the right-hand side of \eqref{cc}, \eqref{re}, \eqref{ccif}
is the smallest for CIF with $q=\infty.$

Clearly, $R^3_{a,q}$ is the smallest for $q=\infty.$ Besides, for each $\beta \in \mathcal{C}_a$ we have $ |\beta|_1 \leq 2|\beta_T|_1/(1-a)$ and $|\beta_T|_1^2 \leq |T| |\beta|_2^2.$ These two  facts imply that $K_a \geq RE_a$ and $\sqrt{RE_a K_a} \leq 2\zeta_{a,2}/(1-a).$ Therefore, we obtain
$$
R^3_{a,\infty} \leq R^3 _{a,2} \leq \frac{2(1+a)|T|^{1/2} \lambda}{
(1-a) \sqrt{RE_a K_a} } \leq \frac{2(1+a)|T|^{1/2} \lambda}{
(1-a) RE_a  }  \leq 2 R^2_a/(1-a).
$$
Taking $a$ not to close to one (for instance, $a=0.5$)  we obtain that \eqref{ccif} with $q=\infty$ is not larger than \eqref{re} with respect to the constant. However, it is possible that $K_a >> RE_a$ \citep{GeerBuhlmann09}, which means that $R^3_{a,\infty}$ might be significantly smaller than $R^2_a.$

Finally, for $\beta \in \mathcal{C} _a$ we have $|\beta|_\infty \leq (1+a)|\beta_T|_1 / (1-a),$ which gives $K_a \leq 2(1+a)|T| \zeta_{a,\infty}/(1-a)^2.$ Consequenlty, $R^3_{a,\infty} \leq (1+a) R^1_a/(1-a),$ so
$R^3_{a,\infty}$ is at most $R^1_{a}$, if one takes $a$ close to zero. Again, we can show the example that   
$R^3_{a,\infty}$ is significanlty larger than $R^1_{a}.$ Consider the orthonormal case $X^TX=\mathbb{I}.$ Then $\zeta_{a,\infty}=1$ and $K_a=\inf_{0 \neq \theta \in \cone _a } \frac{|T||\theta |_2^2}{|\theta_T|_1^2}
\geq 1.$ On the other hand, $K_a$ is smaller than $\frac{|T||\beta|_2^2}{|\beta_T|_1^2}$ for any $\beta \in \mathcal{C}_a.$ Take a vector $d \in \mathbb{R}^p$ such that $d$ has ones on set $T$ and zeroes elsewhere. Clearly, $d \in \mathcal{C}_a$ for any $a.$ Therefore, $K_a \leq \frac{|T||d|_2^2}{|d_T|_1^2}=1,$ so we have $K_a=1.$  Consequently, $R_a^1 =\frac{2(1+a)|T| \lambda}{
(1-a)} ,$ so $R_a^1=\frac{2|T|}{1-a} R^3_{a,\infty} > |T| R^3_{a,\infty} .$ 
Therefore, $R_a^1$ is significantly larger than $R^3_{a,\infty},$ if $|T|$ can tend to infinity.

\bibliographystyle{apalike}
\bibliography{refs}

\end{document}